# Min-Mid-Max Scaling, Limits of Agreement, and Agreement Score

Veli Safak[a]

[a]*Business Administration, Carnegie Mellon University Qatar, Doha, Qatar*

E-mail: vsafak@andrew.cmu.edu

ORCid: 0000-0001-9302-1879

Address: Carnegie Mellon University in Qatar, Education City, Doha, Qatar

# Min-Mid-Max Scaling, Limits of Agreement, and Agreement Score

In this paper, I solve a 60-year old question posed by Cohen's seminal paper (1960) and offer an agreement measure centered around the chance-expected agreement while isolating marginally forced agreement and disagreement. To achieve this, I formulate the minimum feasible agreement given row and column marginals by devising a new algorithm that minimizes the sum of diagonals in contingency tables. Based on this result, I also formulate the lower limit of the most common agreement measure – Cohen's kappa. Finally, I study the lower limit of maximum feasible agreement and devise two statistics of distribution similarity for agreement analysis.

## 1. Introduction

In statistics, a contingency table is a matrix-form table used to display bivariate distributions. Since Pearson (1904), contingency tables are used in several fields to examine the interrelation between raters or methods.

Consider a set of objects on which two raters pass a categorical judgement. The consensus between the raters often measures the reliability of these raters. In this context, the *agreement* is defined as the fraction of observations for which both raters pass the same judgement.

| | **Doctor 2** | |
|---|---|---|
| **Doctor 1** | **Benign** | **Malignant** |
| **Benign** | 0.3 | 0.1 |
| **Malignant** | 0.2 | 0.4 |

**Table 1.** Contingency table of two raters

The contingency table above displays the bivariate distribution of two doctors' diagnoses on some tumor biopsies. In this table, both raters categorize 30% of obser-vations as ‘Benign’ and 40% of observations as ‘Malignant’. The raters agree on 70% of the observations. Is this level of agreement *sufficiently* high?

In the example above, there are two categories: Benign and Malignant. In this paper, I study a general set-up with two raters and $K$ categories: $\{1,2,\ldots,K-1,K\}$. Let $f(i)$ and $g(i)$, respectively, denote the fraction of observations reported in category $i$ by rater-1 and rater-2. Also, let $p(i,j)$ denote the fraction of observations reported in category $i$ by rater-1 and category $j$ by rater-2.

The most celebrated attempt to assess the degree of agreement is Cohen’s kappa proposed by Cohen (1960). Cohen (1960) proposes the chance-expected agreement as a benchmark to assess the degree of agreement. He formulates the chance-expected agreement as $\sum_{i=1}^{K} f(i)g(i)$. The chance-expected agreement corresponds to the agreement level that occurs by chance when raters’ decisions are independent. Cohen (1960) formulates Cohen’s kappa as the fraction of chance-expected disagreement, which does not occur.

$$\kappa = \frac{\sum_{i=1}^{K}\{p(i,i) - f(i)g(i)\}}{1 - \sum_{i=1}^{K} f(i)g(i)}$$

Cohen’s kappa is a commonly used statistic to assess the degree of agreement due to the easy interpretation of its sign. When Cohen's kappa is positive, it indicates that the observed agreement is higher than the chance-expected agreement. Likewise, the negative values of Cohen's kappa indicate that the observed agreement is lower than the chance-expected agreement.

To interpret the magnitude of Cohen’s kappa, researchers had to rely on guidelines that lack theoretical support. The *arbitrary* guidelines proposed by Landis and Koch (1977) and Fleiss (1981) are commonly used and often criticized as they do not respect the marginals. Bakeman et al. (1997) show that the similarities between the raters' marginal probabilities greatly influence the kappa values and conclude that “*no one value of* *kappa can be regarded as universally acceptable."* Similarly, Sims and Wright (2005) find that Cohen's kappa does not sufficiently control for the similarities between the raters' marginal probabilities. Cohen (1960, pg 43) also acknowledges this problem and notes "*disagreement which is forced by marginal disagreement has the same negative consequences as disagreement not so forced-in short, it is disagreement*".

| **Landis and Koch (1977)** | |
|---|---|
| **Kappa Value** | **Interpretation** |
| <0 | No agreement |
| 0-0.20 | Slight |
| 0.21-0.40 | Fair |
| 0.41-0.60 | Moderate |
| 0.61-0.80 | Substantial |
| 0.81-1 | Almost perfect |

| **Fleiss (1981)** | |
|---|---|
| **Kappa Value** | **Interpretation** |
| <0.40 | Poor agreement |
| 0.40-0.75 | Fair/Good agreement |
| >0.75 | Excellent agreement |

**Table 2.** Cohen's kappa magnitude interpretation guidelines

First, let's understand why Cohen's kappa fails. Consider the following two contingency tables with three categories: A, B, and C.

| **(a)** | | **Rater-2** | | |
|---|---|---|---|---|
| | | **A** | **B** | **C** |
| **Rater-1** | **A** | 0 | 0.2 | 0 |
| | **B** | 0.4 | 0 | 0.1 |
| | **C** | 0.2 | 0.1 | 0 |

| **(b)** | | **Rater-2** | | |
|---|---|---|---|---|
| | | **A** | **B** | **C** |
| **Rater-1** | **A** | 0 | 0 | 0.1 |
| | **B** | 0 | 0 | 0.5 |
| | **C** | 0.1 | 0.1 | 0.2 |

**Table 3.** Cohen's kappa and minimum agreement

In the tables above, the chance-expected agreement is 0.3 in Table 3(a) and 0.38 in Table 3(b). The observed agreement is 0 in Table 3(a) and 0.2 in Table 3(b). As a result, Cohen's kappa is $-3/7 \approx -.428$ in Table 3(a) and $-18/62 \approx -.29$ in Table 3(b). Cohen's kappa is higher in Table 3(b). However, it is important to note that the agreement in both tables are at the minimum levels allowed by the marginals. We observe higher agreement level and Cohen's kappa in Table 3(b) because the *marginally forced agreement level* in Table 3(b) is higher. This example shows that Cohen's kappa does not sufficiently control for the marginally forced agreement.

| | **Table 3(a)** | **Table 3(b)** |
|---|---|---|
| **Marginally Forced Agreement (Minimum Agreement)** | 0 | 0.2 |
| **Chance-Expected Agreement** | 0.3 | 0.38 |
| **Observed Agreement** | 0 | 0.2 |
| **Cohen's kappa** | $-3/7 \approx -.428$ | $-18/62 \approx -.29$ |

**Table 4.** Marginally forced agreement and Cohen's kappa

Cohen's kappa does not control for the marginally forced disagreement neither.

| **(a)** | | **Rater-2** A | **Rater-2** B |
|---|---|---|---|
| **Rater-1** | **A** | 0.3 | 0 |
| **Rater-1** | **B** | 0.2 | 0.5 |

| **(b)** | | **Rater-2** A | **Rater-2** B |
|---|---|---|---|
| **Rater-1** | **A** | 0.5 | 0 |
| **Rater-1** | **B** | 0 | 0.5 |

**Table 5.** Cohen's kappa and maximum agreement

In the tables above, the chance-expected agreement is 0.5. The observed agreement is 0.8 in Table 5(a) and 1 in Table 5(b). As a result, Cohen's kappa is 0.6 in Table 5(a) and 1 in Table 5(b). Cohen's kappa is higher in Table 5(b). However, it is important to note that the agreement in both tables are at the maximum levels allowed by the marginals. We observe a higher agreement and Cohen's kappa value in Table 5(b) because the *marginally forced disagreement* is lower in Table 5(b).

| | **Table 5(a)** | **Table 5(b)** |
|---|---|---|
| **Marginally Forced Disagreement (1 – Maximum Agreement)** | 0.2 | 0 |
| **Chance-Expected Agreement** | 0.5 | 0.5 |
| **Observed Agreement** | 0.8 | 1 |
| **Cohen's kappa** | 0.6 | 1 |

**Table 6.** Marginally forced disagreement and Cohen's kappa

As the examples in Table 3 and Table 5 show, Cohen's kappa fails to control for marginally forced agreement and disagreement. This is a major problem limiting the interpretation of Cohen's kappa. Previously, others also noted this problem, but focused on its implications rather than its causes. For example, Warrens (2010) notes "*for fixed observed agreement between the judges, Cohen's kappa penalizes judges with similar marginals compared to judges who produce different marginals*."

We know that the maximum feasible agreement for marginals $f(\cdot)$ and $g(\cdot)$ is formulated as $\sum_{i=1}^{K} min\{f(i), g(i)\}$. However, researchers failed to formulate the minimum feasible agreement in the past six decades. It is equivalent to formulating the lower limit of Cohen's kappa. Cohen (1960) stresses the importance and the complications related to the minimum feasible agreement in agreement analysis. Cohen (1960, pg 41) states "*The lower limit of κ is more complicated, since it depends on the marginal distributions*". Furthermore, Cohen (1960, pg 42) notes "*Since κ is used as a measure of agreement, the complexities of its lower limit are of primarily academic interest*".

All these observations lead us to conclude that an agreement statistic centred around the chance-expected agreement *after isolating the marginally forced agreement and disagreement* can sufficiently measure the *intent-to-agree* between raters. In this paper, I construct this statistic and solve the complex problem of separating the disagreement forced by marginals and disagreement not so forced.

The paper is organized as follows: In Section 2, I formulate the minimum feasible agreement for given marginals, formulate the lower limit of Cohen's kappa, and identify the necessary and sufficient conditions on the marginal under which the maximum feasible agreement is zero. In Section 3, I formulate a piecewise linear scaling technique that centres a variable around a selected value with a range of [0,1]. In Section 4, I apply this scaling technique to the observed agreement level and propose a new agreement statistic centred around the chance-expected agreement after isolating the marginally forced agreement and disagreement. Finally, I study the properties of maximum feasible agreement and propose measures of distribution similarity for agreement analysis in Section 5.

## 2. Limits of Agreement

In this section, I present the general framework and formulate of minimum feasible agreement along with the off-diagonal matching algorithm and the limits of Cohen's kappa.

**Definition 1.** *For given $f(\cdot)$ and $g(\cdot)$, agreement level $A$ is feasible if and only if there exist non-negative $\{p(i,j)\}$ such that*

*(a)* $\sum_{j=1}^{K} p(i,j) = f(i)$ *for all* $i \in \{1,\dots,K\}$,

*(b)* $\sum_{i=1}^{K} p(i,j) = g(j)$ *for all* $j \in \{1,\dots,K\}$, *and*

*(c)* $\sum_{i=1}^{K} p(i,i) = A$.

**Definition 2.** *For given* $f(\cdot)$ and $g(\cdot)$,

*(a) the excess feasible agreement is defined as the difference between the maximum feasible agreement and the chance-expected agreement, and*

*(b) the excess feasible disagreement is defined as the difference between chance-expected agreement and the minimum feasible agreement.*

In this paper, I use the following notation for the critical agreement levels.

| | |
|---|---|
| $A^{obs}$: the observed agreement | $A^{rand}$: the chance-expected agreement |
| $A^{min}$: the minimum feasible agreement | $A^{max}$: the maximum feasible agreement |
| $A^{excess}$: the excess feasible agreement | $D^{excess}$: the excess feasible disagreement |

When the maximum feasible agreement is not 1, it implies that the marginals force a certain degree of disagreement among the raters. This level is called *marginally forced disagreement* and calculated as $1 - A^{max}$. Likewise, when the minimum feasible agreement is not 0, it implies that the marginals force a degree of agreement. In this sense, the minimum feasible agreement represents the *marginally forced agreement*.

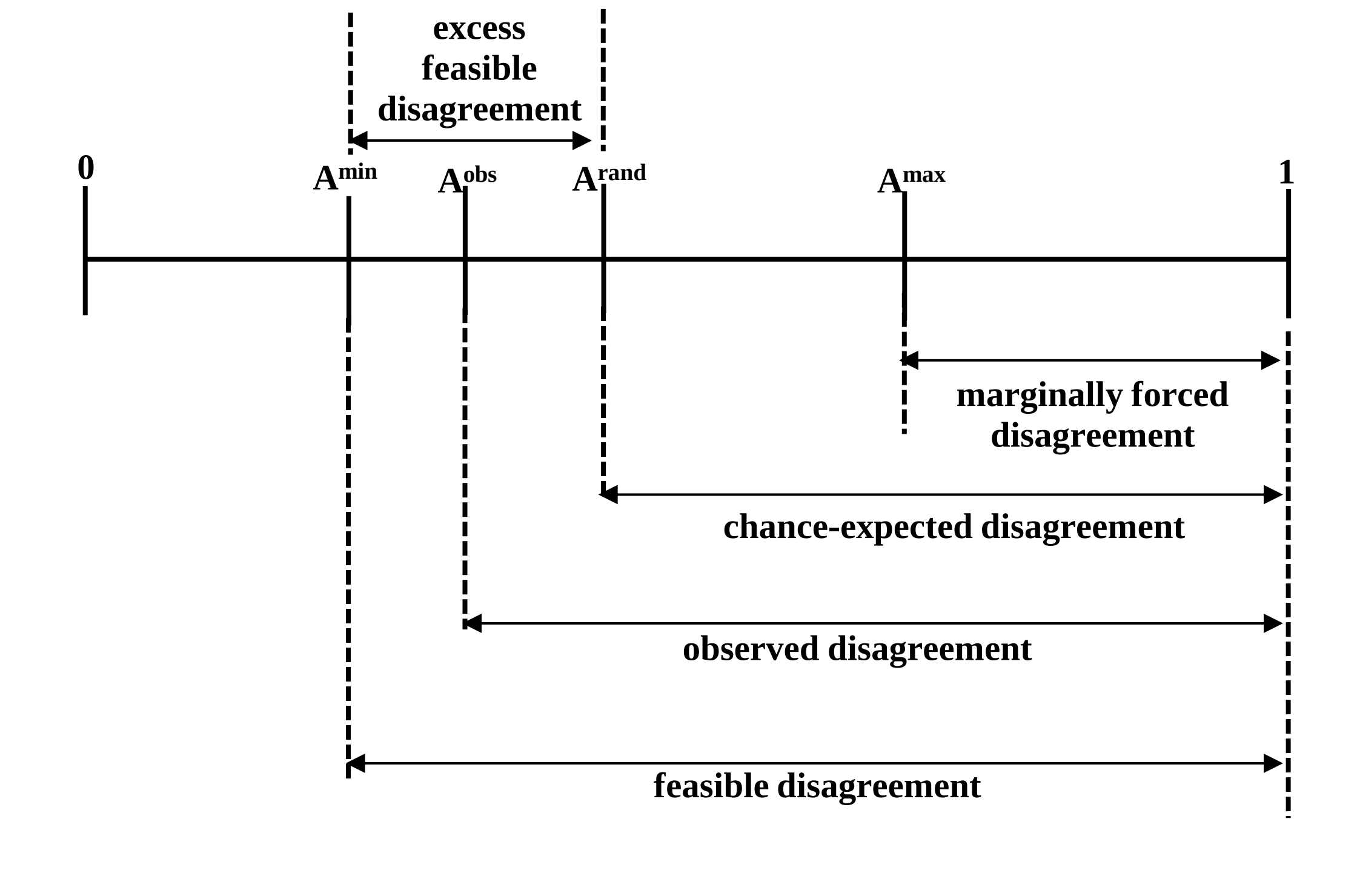

**Figure 1.** Visual representation of key agreement levels for disagreeing raters

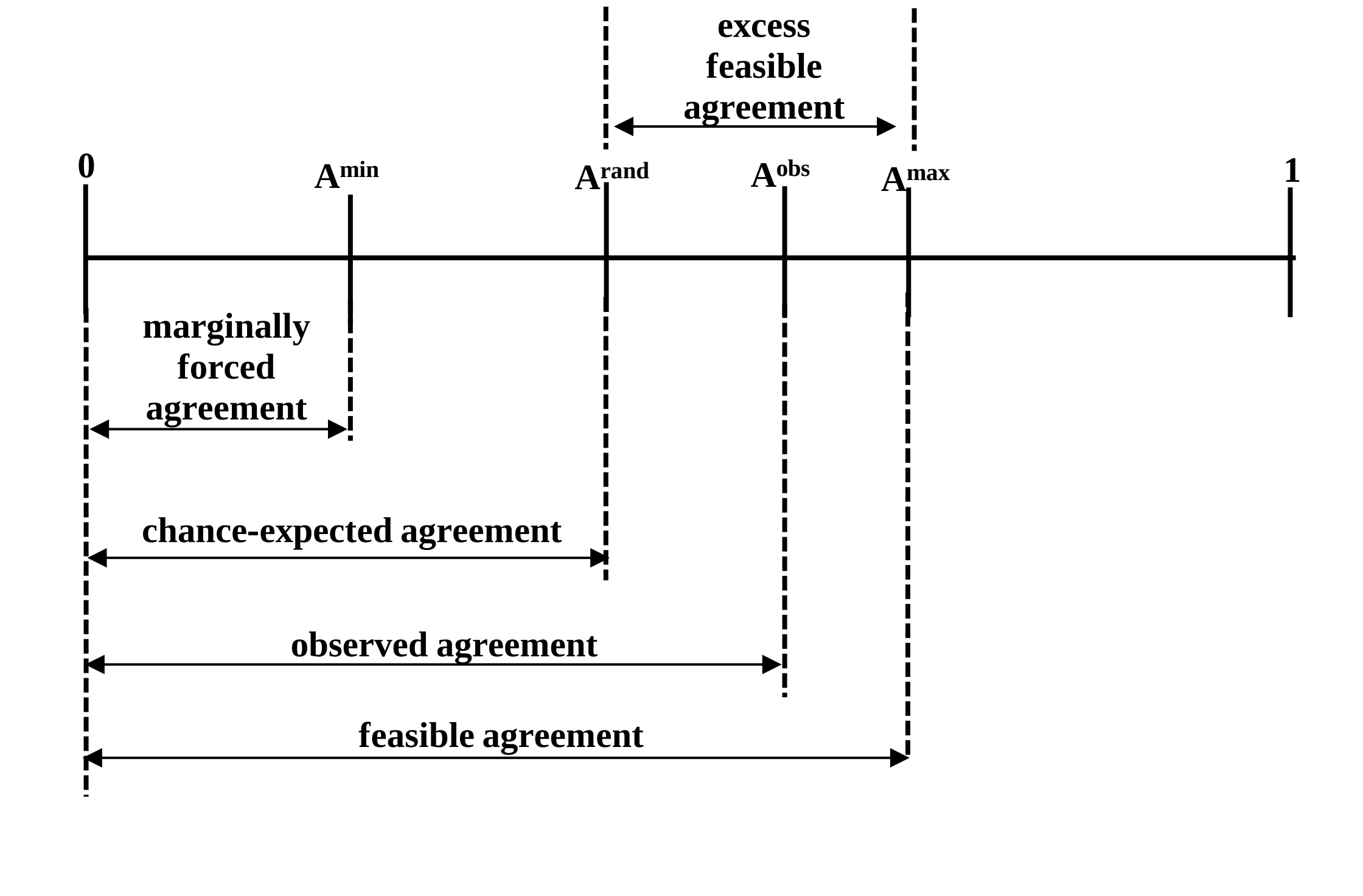

**Figure 2.** Visual representation of key agreement levels for agreeing raters

To formulate the minimum feasible agreement, without loss of generality, I assume that the categories are ordered in a way that the following condition holds.

$$f(i)g(i) \leq f(i+1)g(i+1) \; for \; all \; i \in \{1, \ldots, K-1\} \qquad (1)$$

**Theorem 1.** *The minimum feasible agreement is formulated as follows.*

$$A^{min} = max\{0, f(K) + g(K) - 1\}$$

I prove Theorem 1 by devising an algorithm that generates a contingency table with the minimum feasible agreement for given $f(\cdot)$ and $g(\cdot)$. This algorithm, called *the off-diagonal matching algorithm*, first starts with the random matching as the initial contingency table. Then, it reduces all diagonal cells (up to the last diagonal cell) in a recursive fashion to zero and moves all values in these cells to the neighbouring off-diagonal cells. Regularization condition on Equation (1) makes this operation feasible. Finally, the algorithm looks for rectangular transfers that reduces the last diagonal cell.

To see how this algorithm works, consider the following marginals.

| **Initialization** | **A** | **B** | **C** | **Sum** |
|---|---|---|---|---|
| **A** | 0.03 | 0.02 | 0.05 | 0.1 |
| **B** | 0.09 | 0.06 | 0.15 | 0.3 |
| **C** | 0.18 | 0.12 | 0.3 | 0.6 |
| **Sum** | 0.3 | 0.2 | 0.5 | |

First, we start by creating a table in which the raters decisions are uncorrelated.

| **Step-1 (Iteration-1)** | **A** | **B** | **C** | **Sum** |
|---|---|---|---|---|
| **A** | 0 | 0.05 | 0.05 | 0.1 |
| **B** | 0.12 | 0.03 | 0.15 | 0.3 |
| **C** | 0.18 | 0.12 | 0.3 | 0.6 |
| **Sum** | 0.3 | 0.2 | 0.5 | |

In the first iteration of Step 1, we decrease the cell value of (A,A) to 0 and (B,B) to 0.03 by moving 0.03 from these cells to cells (A,B) and (B,A). This rectangular transfer preserves the row and column marginals.

| Step-1 (Iteration-2) | A | B | C | Sum |
|---|---|---|---|---|
| A | 0 | 0.05 | 0.05 | 0.1 |
| B | 0.12 | 0 | 0.18 | 0.3 |
| C | 0.18 | 0.15 | 0.27 | 0.6 |
| Sum | 0.3 | 0.2 | 0.5 | |

In the next iteration of Step 1, we decrease of (B,B) to 0 and (C,C) to 0.27 by moving 0.03 from these cells to cells (B,C) and (C,B).

| Step-2 (Iteration-1) | A | B | C | Sum |
|---|---|---|---|---|
| A | 0 | 0.05 | 0.05 | 0.1 |
| B | 0 | 0 | 0.3 | 0.3 |
| C | 0.3 | 0.15 | 0.15 | 0.6 |
| Sum | 0.3 | 0.2 | 0.5 | |

The second step looks for rectangular transfers to minimize the remaining non-zero diagonal cell, (C,C). The first iteration of the second step takes place in the lower triangle. We decrease the cell value of (C,C) to 0.15 and (B,A) to 0 by moving 0.12 from these cells to (C,A) and (B,C).

| Step-2 (Iteration-2) | A | B | C | Sum |
|---|---|---|---|---|
| A | 0 | 0 | 0.1 | 0.1 |
| B | 0 | 0 | 0.3 | 0.3 |
| C | 0.3 | 0.2 | 0.1 | 0.6 |
| Sum | 0.3 | 0.2 | 0.5 | |

Once all possible rectangular transfers are exhausted without reducing the last diagonal cell to zero, Step-2 looks for rectangular transfers in the upper diagonal. The second iteration of Step-2 reduces the cell value of (C,C) to 0.1 and (A,B) to 0 by moving 0.05 from these cells to (A,C) and (C,B).

Note that there is no further rectangular transfer to perform in the table above. In this table, the agreement level is 0.1 and it is the lowest possible agreement level. If it was possible to reduce the last diagonal cell to zero before all rectangular transfers were exhausted, then the agreement would be zero.

**Algorithm 1.** The off-diagonal matching algorithm

**Step-0 (Initialization and rectangular transfer function):**

$p(i,j) \leftarrow f(i)g(j)\ for\ all\ i,j \in \{1,\ldots,K\}$

$rectransfer \leftarrow function(i,j,k,l)\{$

$q(i,j,k,l) \leftarrow min\{p(i,j),p(k,l)\}$

$p(i,l) \leftarrow p(i,l) + q(i,j,k,l)$

$p(k,j) \leftarrow p(k,j) + q(i,j,k,l)$

$p(i,j) \leftarrow p(i,j) - q(i,j,k,l)$

$p(k,l) \leftarrow p(k,l) - q(i,j,k,l)$

$\}$

**Step-1 (First update):**

$for\ i\ from\ 1\ to\ K-1\ \{$

do $rectransfer(i,i,i+1,i+1)$

$\}$

**Step-2 (Last update):**

$j \leftarrow 1$

$while\ p(K,K) > 0\ and\ j \leq K-1\{$

$i \leftarrow j+1$

$while\ p(K,K) > 0\ and\ i \leq K-1\{$

do $rectransfer(i,j,K,K)$

$i \leftarrow i+1$

$\}$

$j \leftarrow j+1$

$\}$

$i \leftarrow 1$

$while\ p(K,K) > 0\ and\ i \leq K-1\{$

$j \leftarrow i+1$

$while\ p(K,K) > 0\ and\ j \leq K-1\{$

do $rectransfer(i,j,K,K)$

$j \leftarrow j+1$

$\}$

$i \leftarrow i+1$

$\}$

**Proof of Theorem 1.**

**Lemma 1.** *If* $f(K) + g(K) \leq 1$, *then the off-diagonal matching algorithm creates a contingency table with zero diagonals.*

**Proof of Lemma 1.**

Let *C* denote the total fraction of observations in row *i* and column *j* for all $i, j < K$ after Step-1.

$$C = \sum_{i=1}^{K-1} \sum_{j=1}^{K-1} f(i)g(j) = (1 - f(K))(1 - g(K))$$

Note that until the last iteration in Step-1, *C* does not change. After the last iteration in Step-1, both *C* and $p(K, K)$ decrease by $\sum_{i=1}^{K-1} (-1)^{K-1-i} f(i)g(i)$.

As Step-2 attempts to move the frequency on the last diagonal cell to the off-diagonal cells on the K$^{th}$ row and the K$^{th}$ column, $p(K, K)$ reaches to zero before the total fraction of observations in row *i* and column *j* for all $i, j < K$ if and only if $C \geq f(K)g(K)$. Thus, it suffices to show that $C \geq f(K)g(K)$ when $f(K) + g(K) \leq 1$.

$(1 - f(K))(1 - g(K)) - f(K)g(K)$

$= 1 - f(K) - g(K) + f(K)g(K) - f(K)g(K)$

$= 1 - f(K) - g(K)$

$\geq 0$ ■

The proof of Lemma 1 also shows that we will have a strictly positive last diagonal value while the total fraction of observations in row *i* and column *j* for all $i, j < K$ drops to zero at the end of Step-2 if $f(K) + g(K) > 1$. Under this condition, the minimum feasible agreement level obtains as the remainder in the last diagonal cell at the end of Step-2, i.e., $f(K) + g(K) - 1$. Thus, the minimum feasible agreement obtains as it is stated in Theorem 1. ■■

**Corollary 1.** *For marginals $f(\cdot)$ and $g(\cdot)$, the range of Cohen's kappa is formulated as follows.*

$$\left[\frac{max\{0, f(K)+g(K)-1\}-\sum_{i=1}^{K} f(i)g(i)}{1-\sum_{i=1}^{K} f(i)g(i)}, \frac{\sum_{i=1}^{K}\{min\{f(i), g(i)\}-f(i)g(i)\}}{1-\sum_{i=1}^{K} f(i)g(i)}\right]$$

**Proof.** Immediately follows from Theorem 1.

Next, I identify the necessary and sufficient conditions under which the agreement level can only be zero. I ignore this trivial case in Section 4.

**Proposition 1.** *For marginals $f(\cdot)$ and $g(\cdot)$, $A^{min}=A^{rand}=A^{max}=0$ iff $\{1,\ldots,K\}$ can be partitioned into two sets $I_f$ and $I_g$ such that $\sum_{i\in I_f} f(i)+\sum_{j\in I_g} g(j)=0$.*

**Proof.** By definition, we $0\le A^{min}\le A^{rand}\le A^{max}$. As a result, it suffices to show that $A^{max}=0$ iff $\sum_{i\in I_f} f(i)+\sum_{j\in I_g} g(j)=0$ where $I_f\cup I_g=\{1,\ldots,K\}$.

(⇐) Let $I_f$ *and* $I_g$ *be sets such that* $I_f\cup I_g=\{1,\ldots,K\}$ and $\sum_{i\in I_f} f(i)+\sum_{j\in I_g} g(j)=0$.

$A^{max}=\sum_{i=1}^{K} min\{f(i), g(i)\}$

$=\sum_{i\in I_f} min\{f(i), g(i)\}+\sum_{i\in I_g} min\{f(i), g(i)\}$

$=\sum_{i\in I_f} min\{0, g(i)\}+\sum_{i\in I_g} min\{f(i), 0\}$

$=0+0$

$=0$

(⇒) Suppose that $A^{max}=0$. For this to be true, either $f(i)$ or $g(i)$ must be zero for all $i\in\{1,\ldots,K\}$. Let $I_f:=\{i: f(i)=0\}$ and $I_g:=\{1,\ldots,K\}/I_f$. By construction, we have $\sum_{i\in I_f} f(i)=0$. As $A^{max}=0$, we must also have $\sum_{j\in I_g} g(j)=0$. Thus, we can partition $\{1,\ldots,K\}$ into two set $I_f$ *and* $I_g$ *such that* $\sum_{i\in I_f} f(i)+\sum_{j\in I_g} g(j)=0$ when $A^{max}=0$.■

## 3. Min-Mid-Max Scaling

Feature scaling is an integral part of data pre-processing while working with multiple variables and normalizing statistics. Scaling all features into a common interval reduces the impact of features defined on large scales and allows the small-scale features to contribute equally in optimizing the objective function. Normalizing statistics is equally important to compare their values across different datasets. As the early examples and results show, the range of feasible agreement varies with row and column marginals. For this reason, we need scaling techniques to construct universal measures of agreement.

One of the most common scaling techniques is *min-max scaling*. It is defined as the difference between the feature and its minimum scaled by the difference between its maximum and minimum.

$$f^{min-max}(x) = \frac{x - min}{max - min}$$

Note that the min-max scaling maps random variables to $[0,1]$. However, it does not allow for centring the variable around a selected point, such as mean or median. Centring variables around their means allows us to interpret the transformed variables as deviations from a selected point, such as mean or median. The *mean normalization* centers a feature around its mean, but it fails to scale features on a single range.

$$f^{mean}(x) = \frac{x - mean}{max - min}$$

As both scaling and centering features are useful, I use a scaling technique that can scale and center every feature with compact support.

**Definition 3.** *Consider random variable X defined on support $[a, c]$, and a real number $b$ strictly between $a$ and $c$. The min-mid-max scaling is a mapping defined as follows.*

$$f^{mmm}(x) = \begin{cases} (b-a)^{-1}(x-b) & x \le b \\ (c-b)^{-1}(x-b) & x > b \end{cases}$$

The min-mid-max scaling is a piecewise linear mapping. As a particular case, it is linear when $b = 0.5(a + c)$. Also, it is strictly increasing in x. Thus, it is bijective. Finally, middle point b can be selected by the user. It can be median, mean, or another benchmark. Therefore, min-mid-max scaling offers computational ease and flexibility.

By design, the transformed variable measures deviation of the original variable from b relative to two extremes, minimum and maximum. The transformed variable equals -1 when the original variable is at its minimum, $a$. Also, the transformed variable is 1 when the original variable is at its maximum, $c$. Finally, the transformed variable equals 0 when the original variable equals $b$. As the original variable gets close to $b$, the transformed variable gets close to 0. The sign of the transformed variable indicates the position of the original variable compared to $b$. The transformed variable is negative when the original variable is less than $b$, and positive when the original variable is greater than $b$.

## 4. Agreement Score

In this section, I consider non-trivial cases that do not satisfy the premise of Corollary 2. I start by proposing a measure of inter-rater agreement that eliminates marginally forced disagreement and agreement by using min-max scaling the observed agreement level.

**Definition 4.** *For given* $f(\cdot)$ and $g(\cdot)$, *the agreement score, denoted by* $S$, *is calculated as follows.*

$$S = \frac{A^{obs} - A^{min}}{A^{max} - A^{min}}$$

It is important to note that the agreement score is between zero and one. As the observed agreement gets closer to the maximum feasible agreement, the agreement score gets closer to one. Likewise, an agreement score close to zero indicates that the observed agreement is close to the minimum feasible agreement. These limits remove the marginally force agreement and disagreement from the calculation.

Although the agreement score allow us to establish the position of the observed agreement level with respect to the minimum and maximum feasible agreement levels, it does not allow us to sufficiently identify the cases where raters are in disagreement more than agreement. As Cohen (1960) proposes, the chance-expected agreement serves as a good benchmark above which the raters are considered to be in agreement and below which the raters are considered to be in disagreement.

To incorporate the chance-expected agreement into the calculations, I apply min-mid-max scaling to the observed agreement level with the minimum feasible agreement, the chance-expected agreement, and the maximum feasible agreement. I define this measure as *the centralized agreement score*.

**Definition 5.** *For given* $f(\cdot)$ and $g(\cdot)$, *the centralized agreement score, denoted by* $\mathbb{S}$, *is calculated as follows.*

$$\mathbb{S} = \begin{cases} -\dfrac{A^{rand} - A^{obs}}{A^{rand} - A^{min}} & A^{obs} \leq A^{rand} \\ \dfrac{A^{obs} - A^{rand}}{A^{max} - A^{rand}} & A^{obs} > A^{rand} \end{cases}$$

First of all, notice that the sign of the agreement score indicates the position of the observed agreement level relative to the chance-expected agreement. When the observed agreement level is lower than the chance-expected agreement, then the agreement score is negative. When the observed agreement level is higher than the chance-expected agreement, then the agreement score is positive. Finally, the agreement score is zero when observed agreement level equals to the chance-expected agreement. In this perspective, the sign of agreement score shares the same interpretation as Cohen's kappa.

To interpret the absolute value of the agreement score, remember that the excess feasible disagreement is defined as the difference between the chance-expected agreement and the minimum feasible agreement, i.e., $A^{rand} - A^{min}$. As a result, the absolute value of the agreement score can be interpreted as the share of realized excess feasible disagreement when the observed agreement is lower than the chance-expected agreement.

Likewise, I define the excess feasible agreement as the difference between the maximum feasible agreement and the chance-expected agreement, i.e., $A^{max} - A^{rand}$. Thus, the absolute value of the agreement score can be interpreted as the share of realized excess feasible agreement when the observed agreement level is greater than the chance-expected agreement.

Since the centralized agreement score measures the share of excess feasible agreement/disagreement that realizes, it eliminates the marginally forced agreement and disagreement levels. For this reason, I propose it as a measure of *intent-to-agree*. As it gets closer to -1, it indicates lower intent-to-agree. On the other hand, values closer to 1 indicates high degree of intent-to-agree. The centralized agreement score can be used as a measure of observed agreement to compare multiple contingency table only when they have the same range of feasible agreement and the chance-expected agreement.

It is also easy to map the centralized agreement score to actual agreement values. Consider the following contingency tables.

**(a)**

| | | **Rater-2** | | |
|---|---|---|---|---|
| | | **A** | **B** | **Total** |
| **Rater-1** | **A** | 0.1 | 0 | 0.1 |
| | **B** | 0.8 | 0.1 | 0.9 |
| | **Total** | 0.9 | 0.1 | |

**(b)**

| | | **Rater-2** | | |
|---|---|---|---|---|
| | | **A** | **B** | **Total** |
| **Rater-1** | **A** | 0.2 | 0.3 | 0.5 |
| | **B** | 0.3 | 0.2 | 0.5 |
| | **Total** | 0.5 | 0.5 | |

**Table 7.** Marginal probabilities and intent-to-agree

Note that the centralized agreement score in Table 7(a) and Table 7(b) are, respectively, 1 and -0.2. Although the raters in Table 7(a) want to agree as much as possible, the differences in their biases towards the two classes (A and B) do not allow them to agree on more than 20% of the observations.

While the raters in Table 7(b) do not wish to agree as much as the raters in Table 7(a), they agree on 40% of the cases because of the similarity between their marginal distributions. What would be the agreement level if raters in Table 7(a) rated each option with equal likelihood like the raters in Table 7(b)? In this case, they would agree on all cases with perfect intent-to-agree. I call this hypothetical agreement level as *no-bias agreement* and formulate it as follows.

**Definition 6.** *Consider a contingency table with $K > 0$ number of classes for which $f(i) = g(j) = 1/K$ for all $i, j \in \{1, \dots, K\}$. Given a level of centralized agreement score $\mathbb{S}$, no-bias agreement, denoted by $A^{NB}$, is defined as the level of agreement associated with centralized agreement score $\mathbb{S}$ that would occur when raters rate each option with equal likelihood and formulated as follows.*

$$A^{NB} = \begin{cases} \dfrac{\mathbb{S} + 1}{K} & \mathbb{S} \leq 0 \\ \dfrac{(K-1)\mathbb{S} + 1}{K} & \mathbb{S} > 0 \end{cases}$$

The no-bias agreement level is the agreement level that corresponds to the agreement level that is observed with the centralized agreement score $\mathbb{S}$ when row and column marginal distributions are uniform. This agreement statistic allows us to map different degrees of intent-to-agree observed in contingency tables with different marginals into agreement levels in a standard contingency table with uniform row and column marginals. As this transformation is bijection, one can use the hypothetical no-bias agreement levels and the centralized agreement scores to compare the degree of intent-to-agree across contingency tables.

In addition to the uniform normalization, we can map a centralized agreement score calculated in a contingency table into the agreement level that we would observe in a contingency table with different number of classes and different marginals.

Let $\{f^{[1]}(\cdot), g^{[1]}(\cdot)\}$ be row and column marginals for a contingency table of M classes. Likewise, let $\{f^{[2]}(\cdot), g^{[2]}(\cdot)\}$ be row and column marginals for a contingency table of K classes. Assume that both of these marginal combinations satisfy the premise of Theorem 1. More specifically, suppose that we have

a) $f^{[1]}(i)g^{[1]}(i) \leq f^{[1]}(i+1)g^{[1]}(i+1)$ for all $i \in \{1, \ldots, M-1\}$, and

b) $f^{[2]}(i)g^{[2]}(i) \leq f^{[2]}(i+1)g^{[2]}(i+1)$ for all $i \in \{1, \ldots, K-1\}$.

**Definition 7.** *Consider a contingency table with marginals* $f^{[1]}(\cdot)$ *and* $g^{[1]}(\cdot)$. *Given a level of centralized agreement score* $\mathbb{S}$, *hypothetical agreement with marginals* $f^{[2]}(\cdot)$ *and* $g^{[2]}(\cdot)$ , *denoted by* $A^H(f^{[2]}, g^{[2]})$, *is defined as the level of agreement associated with centralized agreement score* $\mathbb{S}$ *that would occur when raters have marginals* $f^{[2]}(\cdot)$ *and* $g^{[2]}(\cdot)$ *and formulated as follows*

$$A^H(f^{[2]}, g^{[2]}) = \begin{cases} (\mathbb{S}+1)A^{rand}(f^{[2]}, g^{[2]}) - \mathbb{S}A^{min}(f^{[2]}, g^{[2]}) & \mathbb{S} \leq 0 \\ \mathbb{S}A^{max}(f^{[2]}, g^{[2]}) + (1-\mathbb{S})A^{rand}(f^{[2]}, g^{[2]}) & \mathbb{S} > 0 \end{cases}$$

*where*

$$A^{rand}(f^{[2]}, g^{[2]}) = \sum_{i=1}^{K} f^{[2]}(i)g^{[2]}(i),$$

$$A^{min}(f^{[2]}, g^{[2]}) = max\{0, f^{[2]}(K) + g^{[2]}(K) - 1\}, and$$

$$A^{max}(f^{[2]}, g^{[2]}) = \sum_{i=1}^{K} min\{f^{[2]}(i), g^{[2]}(i)\}$$

It is important to note that the hypothetical agreement proposed in Definition 7 does not require the actual marginals and the hypothetical marginals to share the same number of classes. As such, it allows comparisons of intent-to-agree across a broader range of contingency tables. This feature is particularly useful to compare intent-to-agree across contingency tables in which the same variable is categorized in different classes, such as credit ratings proposed by different credit rating agencies. Like the no-bias agreement, the hypothetical agreement with marginals $f^{[2]}(\cdot)$ and $g^{[2]}(\cdot)$ is a bijection of the centralized agreement score. Thus, it highlights the applicability of the centralized agreement score for intent-to-agree comparison across a large set of contingency tables.

## 5. Distribution Similarity

The results presented in the previous sections highlight the importance of the similarity between row and column marginal distributions. In this section, I propose a measure of the distribution similarity by using the maximum feasible agreement. All the results in this section apply to the trivial case presented in Corollary 2. Recall that the maximum feasible agreement is denoted by $A^{max}$ and formulated as follows.

$$A^{max} = \sum_{i=1}^{K} min\{f(i), g(i)\}$$

It is important to note that the maximum feasible agreement can reach 1 if and only if the raters have the same distribution. This observation is the first reason why the maximum feasible agreement is a good candidate to measure the distribution similarity. The second reason is the symmetry of the maximum feasible agreement. Symmetry is intuitively an important attribute for similarity measures. If $f(\cdot)$ is similar to $g(\cdot)$ with a certain degree, then $g(\cdot)$ must be similar to $f(\cdot)$ with the same degree.

First, I present the range of the maximum feasible agreement in Theorem 2. The results presented in Theorem 2 allow me to calculate a relative similarity score of distribution $g(\cdot)$ for distribution $f(\cdot)$ by comparing the similarity between $g(\cdot)$ and $f(\cdot)$ to the similarity between $f(\cdot)$ and all other distributions.

**Theorem 2.** *Let $f(\cdot)$ be a distribution such that $f(i^*) \leq f(i)$ for some $i^* \in \{1, \dots, K\}$ and for all $i \in \{1, \dots, K\}$. The maximum feasible agreement between $f(\cdot)$ and distribution $g(\cdot)$*

*(a) ranges between $f(i^*)$ and one,*

*(b) is zero if and only if $f(i^*) = 0$, $g(i) = 0$ for all $i \in \{1, \dots, K\}/\{argmin_i f(i)\}$ and $\sum_{i^* \in argmin_i f(i)} g(i^*) = 1$, and*

*(c) is strictly positive and equals to $f(i^*)$ if and only if $f(i^*) > 0$, $g(i^*) = 1$ for some $i^* \in \{argmin_i f(i)\}$, $g(i) = 0$ for all $i \in \{1, \dots, K\}/\{argmin_i f(i)\}$.*

**Proof**

Without loss of generality, assume that $f(1) \leq f(2) \leq \cdots \leq f(K)$.

Let $j$ be the smallest index such that $f(i) > 0$ for all $i \in \{j, \dots, K\}$.

**Case 1:** $j = 1$, $g(1) = 1$ and $g(i) = 0$ *for all* $i \in \{2, \dots, K\}$

$$A^{max} = \sum_{i=1}^{K} min\{f(i), g(i)\} = min\{f(1), g(1)\} + \sum_{i=2}^{K} min\{f(i), g(i)\}$$

$$= min\{f(1), 1\} + \sum_{i=2}^{K} min\{f(i), 0\} = f(1)$$

**Case 2:** $j = 1$ and $f(1) < g(1) < 1$

$$A^{max} = \sum_{i=1}^{K} min\{f(i), g(i)\}$$

$$= min\{f(1), g(1)\} + \sum_{i=2}^{K} min\{f(i), g(i)\}$$

$$> f(1)$$

The last inequality obtains because there exists at least one strictly positive term in the summation as $f(i) > 0$ for all $i \in \{1, \dots, K\}$ and $\sum_{i=2}^{K} g(i) > 0$.

**Case 3:** $j \neq 1, \sum_{i=1}^{j-1} g(i) = 1$ and $g(i) = 0$ *for all* $i \in \{j, \dots, K\}$

$$A^{max} = \sum_{i=1}^{K} min\{f(i), g(i)\}$$

$$= \sum_{i=1}^{j-1} min\{f(i), g(i)\} + \sum_{i=j}^{K} min\{f(i), g(i)\}$$

$$= \sum_{i=1}^{j-1} min\{0, g(i)\} + \sum_{i=j}^{K} min\{f(i), 0\}$$

$$= 0$$

**Case 4:** $j \neq 1, \sum_{i=1}^{j-1} g(i) < 1$

$$A^{max} = \sum_{i=1}^{K} min\{f(i), g(i)\}$$

$$= \sum_{i=1}^{j-1} min\{f(i), g(i)\} + \sum_{i=j}^{K} min\{f(i), g(i)\}$$

$$= 0 + \sum_{i=j}^{K} min\{f(i), g(i)\}$$

$$> 0$$

The last inequality obtains because there exists at least one strictly positive term in the summation because $f(i) > 0$ for all $i \in \{j, \dots, K\}$ and $\sum_{i=j}^{K} g(i) > 0$. ■

Theorem 2 establishes that the maximum feasible agreement for distribution $f(\cdot)$ and all other distributions ranges between $min_i f(i)$ and one. I propose using the min-max scaled version of the maximum feasible agreement between $g(\cdot)$ and $f(\cdot)$ to measure the *agreement-similarity of distribution* $g(\cdot)$ *to distribution* $f(\cdot)$.

**Definition 8.** *For given* $f(\cdot)$ and $g(\cdot)$, *the agreement-similarity score of distribution* $g(\cdot)$ *to distribution* $f(\cdot)$, *is formulated as follows.*

$$AS(g, f) = \frac{\left(\sum_{i=1}^{K} min\{f(i), g(i)\}\right) - min_i f(i)}{1 - min_i f(i)}$$

Note that the agreement-similarity score ranges between zero and one. As it is close to one, it indicates a high agreement-similarity between $f(\cdot)$ and $g(\cdot)$ because the maximum feasible agreement also is also close to one. On the other hand, it is close to zero when the agreement-similarity is low because the maximum feasible agreement given $f(\cdot)$ is near its lower limit.

## 6. Conclusion

In this paper, I contribute to the literature on contingency tables in four-folds.

First, I present the off-diagonal matching algorithm that minimizes the sum of diagonal cells in contingency tables for given marginal distributions. The algorithm converges $2K^2 - 2K + 1$ iterations at most for contingency tables with $K$ classes.

Secondly, I formulate the minimum feasible agreement for given marginal distributions by using the off-diagonal matching algorithm. Based on this result, I formulate the lower limit of the most famous agreement measure, Cohen's kappa. This result allows the researchers to interpret Cohen's kappa values in a more meaningful way backed by theoretical foundation.

Thirdly, I offer two new measures of agreement: agreement score and centralized agreement score. The first one that can be used to assess the degree of agreement within the range of feasible agreement for given marginals. The second one allows us to measure the intent-to-agree between two raters. By constructing two hypothetical agreement levels based on the intent-to-agree for different row and column marginals, I show that the centralized agreement score can be mapped into agreement space and be used to compare intent-to-agree across contingency tables with different marginals and number of classes.

Finally, I show that the maximum feasible agreement is a good measure of similarity between two distributions for agreement analysis. I formulate the lower limit of the maximum feasible agreement and present a measure of similarity between two distributions by using one of them as the base distribution and assessing the similarity of the other relative to all distributions.

Overall, this paper provides a solution for a long-standing problem in contingency table literature and offer new insights on the inter-rater agreement.